\documentstyle{llncs}

\begin{document}

\title{Fault-Tolerant Quantum Computation with Higher-Dimensional Systems}

\author{Daniel Gottesman\thanks{gottesma@t6-serv.lanl.gov}}

\institute{T-6 Group, Los Alamos National Laboratory}

\maketitle

\begin{abstract}
Instead of a quantum computer where the fundamental units are
2-dimensional qubits, we can consider a quantum computer made up of
$d$-dimensional systems.  There is a straightforward generalization of
the class of stabilizer codes to $d$-dimensional systems, and I will
discuss the theory of fault-tolerant computation using such codes.  I
prove that universal fault-tolerant computation is possible with any
higher-dimensional stabilizer code for prime $d$.
\end{abstract}

\section{Introduction}

Quantum computation and quantum communications have the potential to
accomplish many things that would be difficult or impossible using
just classical computers and communications.  However, quantum data is
very vulnerable to decoherence and to errors.  It is likely that some
form of quantum error correction will be needed to perform anything
beyond the simplest computations with a quantum computer.  Quantum
error-correcting codes~\cite{shor9,steane7,stabilizer,GF4a,GF4b}
provide one of the tools necessary.  Such a code can protect quantum
data against errors occuring during transmission or storage of the
data.  However, to have a reliable quantum computer, we also need for
the computation to be performed in a fault-tolerant
manner~\cite{shorFT}.  Fault-tolerant quantum computation requires a
protocol that not only maps states of a quantum code to other states
of a quantum code, but prevents errors from propagating out of
control.

A large group of useful codes was introduced in \cite{stabilizer} and
\cite{GF4a}.  These codes are essentially the quantum equivalent of
classical linear codes, in that they can be easily described and
encoded, and that it is easy to measure the error syndrome (as in the
classical case, it may be difficult to compute the actual error from
the error syndrome).  The primary complication involved in quantum
error correction is that it is not only necessary to correct bit flip
errors
\begin{equation}
X = \pmatrix{0 & 1 \cr 1 & 0} \enspace,
\end{equation}
but also phase errors
\begin{equation}
Z = \pmatrix{1 & 0 \cr 0 & -1} \enspace.
\end{equation}
Consequently, it turns out to be useful to look at the group $\cal{P}$
generated by tensor products of these operators.  $\cal{P}$ is called
the Pauli group or the extra-special group.  {\it Stabilizer codes}
are those codes where the valid codewords are all eigenstates of $n-k$
operators in $\cal{P}$.  These $n-k$ operators generate a $2^{n-k}$
element Abelian group, called the {\it stabilizer} $S$ of the code.
The set of valid codewords forms a $2^k$-dimensional subspace of the
full $n$-qubit Hilbert space, the {\it coding space} of the code.  The
stabilizer is analogous to the parity check matrix of a classical
linear code.  In fact, the set of classical linear binary codes is
exactly the set of stabilizer codes where everything in the stabilizer
is a tensor product of $Z$ operators.

Stabilizer codes are easy to work with because of the structure of the
Pauli group.
\begin{equation}
XZ = - ZX \enspace,
\end{equation}
so any two operators in $\cal{P}$ either commute or they anticommute.
If an operator $E$ anticommutes with an operator $G$ in the stabilizer
$S$ of a code, then $E |\psi\rangle$ will have eigenvalue $-1$ for $G$
instead of $+1$.  Therefore, by measuring the eigenvalues of the $n-k$
generators of $S$, we can easily measure the error syndrome, and if
the code is suitably chosen, identify the error.

Even if we restrict our attention to stabilizer codes, it is far from
obvious that we can perform fault-tolerant computation.  First of all,
we must find some operators that map the coding space of the code into
itself.  Secondly, many of these operators will cause errors to spread
from one qubit in a block to a different qubit in the same block of
the code.  Therefore, a single error could rapidly grow to become many
errors within a block, exceeding the code's capability to correct
them.  In order to avoid this, we will further restrict attention to
{\it transversal} operations, that is, operations which only interact
qubits from one block with corresponding qubits in other blocks.  This
means that an error occuring in qubit number $3$ in one block might
spread to qubit number $3$ in another block, but it will never spread
back to qubit $2$ in the first block.  Each block can easily correct a
single error, so this situation causes no problems.  Shor was the
first to demonstrate a universal set of fault-tolerant
gates~\cite{shorFT}.  However, his construction only worked for a
small class of codes.  In~\cite{faulttol}, I was able to show that
Shor's construction could be generalized to any qubit stabilizer code.
The proof made extensive use of the group of unitary operators that
leave the group ${\cal P}$ invariant under conjugation.  This group is
known as the Clifford group.

In the classical theory of error-correcting codes, it is often helpful
to go beyond bits and work with higher-dimensional systems.  The same
may be true for quantum error correction.  Instead of a system made up
of 2-dimensional qubits, we can work with a system composed of
$d$-dimensional {\it qudits}.  It turns out that there is a natural
generalization of stabilizer codes to higher-dimensional systems
\cite{knill1,knill2,rains}.  In Sec.~\ref{qudits}, I will present this
generalization.  Then I will procede to generalize the arguments of
\cite{faulttol} to show that universal fault-tolerant computation is
also possible with any of these codes, at least in the case where $d$
is prime.  Though I will largely focus on prime $d$, I will also say a
little bit about the general case along the way.  Assume that $d$ is
prime unless it is otherwise specified.

The construction of a universal set of gates used in \cite{faulttol}
consisted of a number of steps:
\begin{enumerate}
\item \label{Clifford} The full Clifford group can be constructed
given the controlled-NOT, operators in the Pauli group, and
measurement of operators in the Pauli group.
\item \label{Pauli} For any stabilizer code, we can perform encoded
versions of all operators in the Pauli group and can measure operators
in the Pauli group.
\item \label{CNOT} We can perform a CNOT between corresponding encoded
qubits in different blocks of the code.
\item \label{swap} We can swap individual encoded qubits from one
block of the code to an empty block and vice-versa.
\item \label{move} We can move an encoded qubit in an otherwise empty
block to whatever position in the block we desire.  At this point, we
have established that we can perform a CNOT between any pair of
encoded qubits in the same or different blocks, and therefore can
perform the full Clifford group.
\item \label{universal} Given the full Clifford group, we can perform
an additional gate outside the Clifford group, such as the Toffoli
gate or the $\pi/8$ rotation.  This completes the universal set of
gates.
\end{enumerate}
Each of the steps in this construction has an analog for
higher-dimensional systems.  However, in this paper, I will present a
simplified construction that combines steps \ref{CNOT} through
\ref{move}.  This construction gives a direct way to perform the
generalization of the CNOT between any pair of encoded qudits, whether
they are in the same or different blocks and whether they are in
corresponding or different positions within their blocks.  This
construction can also be used to simplify the proof for systems with
$d=2$, and will likely reduce the required overhead for fault-tolerant
computation using codes with many qudits per block.

\section{Higher Dimensional Generalization of the Pauli
Group and Other Structures}
\label{qudits}

The Pauli group has a natural generalization to higher-dimensional
systems.\footnote{The group presented in this section is not the only
generalization of the Pauli group, although it is probably the
simplest.  See \cite{knill1} for a more extensive discussion of this
issue.}  Instead of generating it from the two-dimensional $X$ and
$Z$, we instead generate ${\cal P}$ from tensor products of $X_d$ and
$Z_d$, where $X_d |j\rangle = |j+1\rangle$ and $Z_d |j\rangle =
\omega^j |j\rangle$, where $\omega$ is a primitive $d$-th root of
unity.  $X_d$ and $Z_d$ satisfy the relation
\begin{equation}
X_d Z_d = \omega^{-1} Z_d X_d \enspace.
\end{equation}
Both $X_d$ and $Z_d$ have order $d$.  The elements of the single-qudit
Pauli group have the form $\omega^a X_d^r Z_d^s$, where $0 \leq r, s <
d$, and
\begin{equation}
\left( X_d^r Z_d^s \right) \left( X_d^t Z_d^u \right) = \omega^{st-ru}
\left( X_d^t Z_d^u \right) \left( X_d^r Z_d^s \right) \enspace.
\end{equation}
${\cal P}$ for $n$ qubits will contain $d^{2n}$ elements, plus an
additional factor of $d$ for overall phase.  The elements of ${\cal
P}$ have eigenvalues $\omega^r$ for $r = 0, \ldots,
d-1$.\footnote{This is true for odd $d$.  For even $d$, $XZ$ has
order $2d$, so extra factors of $i$ will be necessary, as in the $d=2$
case.  This aspect is actually simpler for odd $d$ than for $d=2$.}
From now on, I will suppress the subscript $d$, and all operations
should be taken to be over qudits instead of qubits.

The $d$-dimensional generalization of the stabilizer $S$ of a code is
again just an Abelian subgroup of ${\cal P}$.  The coding space is
composed of those states that are fixed by all elements of $S$ (when
$d$ is even, this actually imposes an additional constraint on the
overall phase of elements of $S$).  If the stabilizer on $n$ qudits
has $n-k$ generators, then $S$ will have $d^{n-k}$ elements and the
coding space will consist of $k$ qudits.  Note that this last fact
need no longer be true when $d$ is not prime, and this is the main
source of complications in that case.  It is unclear exactly how to
deal with a code that does not encode an integral number of qudits.
If we stick to codes for which all the generators of the stabilizer
have order $d$, the rest of the proof will hold, modulo a question
about gates necessary to generate the Clifford group.

If an operator $E$ and $M \in S$ satisfy
\begin{equation}
E M = \omega^a M E \enspace,
\end{equation}
then $E |\psi\rangle$ will have eigenvalue $\omega^a$ for $M$ instead
of eigenvalue $+1$, so we can detect that error $E$ has occurred by
measuring the eigenvalue of $M$.

We can see the structure of the coding space by extending the
generators of $S$ to a complete independent set of commuting
operators.  When $d$ is not prime, we also require these operators to
have order $d$.  Such a set will have cardinality $n$, so we can do
this by choosing $k$ additional operators $\overline{Z}_1, \ldots,
\overline{Z}_k$.  These operators have the interpretation of the
encoded $Z$ operators for the $k$ encoded qudits.  We can then choose
$k$ more operators $\overline{X}_1, \ldots, \overline{X}_k$ which
satisfy the relations
\begin{eqnarray}
\overline{X}_i \overline{Z}_j & = & \overline{Z}_j \overline{X}_i \ (i
\neq j) \\
\overline{X}_i \overline{Z}_i & = & \omega^{-1} \overline{Z}_i
\overline{X}_i \\
\overline{X}_i M & = & M \overline{X}_i \ (\forall M \in {\cal P}) \enspace.
\end{eqnarray}
The operators $\overline{X}_i$ then act as the encoded $X$ operators
for the $k$ encoded qudits.  The generators of $S$ along with
$\overline{Z}_1, \ldots, \overline{Z}_k$ and $\overline{X}_1, \ldots,
\overline{X}_k$ then generate the group of all Pauli group operators
that commute with $S$.  As in the two-dimensional case, the operators
that commute with $S$ but are not themselves in $S$ are precisely the
operators that cannot be detected by the quantum code.  They therefore
perform encoded operations on the data.

The Clifford group is the set of operators that leave ${\cal P}$
invariant under conjugation.  That is, it is the normalizer $N({\cal
P})$ of ${\cal P}$ in the unitary group $U(d^n)$.  The Clifford group
is important for fault-tolerant computation because if we perform an
operator $U$ on the Hilbert space, the operator $U N U^\dagger$ has
the same relationships to states after the transformation as the
operator $N$ did before the transformation.  Therefore, instead of
considering transformations of the states $|\psi\rangle \rightarrow U
|\psi\rangle$, we can consider transformations of the operators $N
\rightarrow U N U^\dagger$.  When $U$ is in the Clifford group and $N$
is in the Pauli group, then $U N U^\dagger$ will also be in the Pauli
group.  Therefore, we can uniquely describe elements of $N({\cal P})$
by the permutation they induce on ${\cal P}$.  The permutation must
preserve the group structure of ${\cal P}$, but is otherwise
arbitrary.

In the two-dimensional group, $N({\cal P})$ was generated by two
single-qubit operators $R$ (the Hadamard transform $|j\rangle
\rightarrow |0\rangle + (-1)^j |1\rangle$) and $P$ (the phase gate
$|j\rangle \rightarrow i^j |j\rangle$), and the two-qubit operator CNOT
($|i\rangle |j\rangle \rightarrow |i\rangle |(i + j) \bmod 2
\rangle$).  In $d$ dimensions, $R$ generalizes to the $d$-dimensional
discrete Fourier transform
\begin{equation}
|j\rangle \rightarrow \sum_{s=0}^d \omega^{js} |s\rangle \enspace,
\end{equation}
$P$ generalizes to the $d$-dimensional phase gate
\begin{equation}
|j\rangle \rightarrow \omega^{j(j-1)/2} |j\rangle \enspace,
\end{equation}
and CNOT generalizes to the SUM gate
\begin{equation}
|i\rangle |j\rangle \rightarrow |i\rangle |(i + j) \bmod d \rangle
\enspace.
\end{equation}

We can describe these operators by their induced transformations on
the Pauli group.  $R$ maps
\begin{eqnarray}
X & \rightarrow & Z \enspace, \\
Z & \rightarrow & X^{-1} \enspace.
\end{eqnarray}
$P$ maps
\begin{eqnarray}
X & \rightarrow & XZ \enspace, \\
Z & \rightarrow & Z \enspace.
\end{eqnarray}
SUM maps
\begin{eqnarray}
X \otimes I & \rightarrow & X \otimes X \enspace, \\
I \otimes X & \rightarrow & I \otimes X \enspace, \\
Z \otimes I & \rightarrow & Z \otimes I \enspace, \\
I \otimes Z & \rightarrow & Z^{-1} \otimes Z \enspace.
\end{eqnarray}
However, it is not clear that these three gates generate the Clifford
group.  We may also need the $S$ gate
\begin{eqnarray}
X & \rightarrow & X^a \enspace, \\
Z & \rightarrow & Z^b \enspace,
\end{eqnarray}
for all pairs $(a,b)$, where $ab \equiv 1 \bmod d$.  On kets, this
gate acts as $|j\rangle \rightarrow |aj\rangle$.  In fact, a single
pair $(a,b)$ is sufficient, as long as $a$ generates the
multiplicative group $\bbbz_d^*$.  I will not give a detailed proof
that these gates generate the Clifford group, but using the $P$, $R$,
and $S$ gates, we can get the full one-qudit Clifford group.  Then a
construction similar to that used in \cite{faulttol} will give the
full $n$-qudit Clifford group.  The structure is somewhat more
complicated when $d$ is not prime, and I have not verified that these
gates are sufficient for the nonprime case.

Note that we can fault-tolerantly measure any operator in ${\cal P}$
that is the tensor product of $Z$ operators by performing transversal
SUM gates from the qudits to be measured to an appropriate ancilla.
Since we are interested in eigenvalues with possible values $\omega^j$
for $j=0, \ldots, d-1$, the appropriate ancilla state is the
superposition of all states where the sum of the qudits is $0 \bmod
d$.  That way, no information beyond the eigenvalue of the measured
operator will be conveyed.  We can construct this state by Fourier
transforming the state $\sum_{j=0}^{d-1} |j j \cdots j \rangle$.
Following DiVincenzo and Shor~\cite{divince}, we can measure any
operator in ${\cal P}$ by performing a transversal Clifford group
operation $C$ that takes the operator to the tensor product of $Z$'s,
performing the measurement, and applying $C^{-1}$.

For any stabilizer code, the elements of the $k$-qudit encoded Pauli
group are also elements of the $n$-qudit unencoded Pauli group, as are
the generators of the stabilizer.  Since we can perform and measure an
arbitrary element of the unencoded Pauli group, we have shown that for
a stabilizer code over $d$-dimensions, we can apply encoded versions of
$X$ and $Z$ for all encoded qudits, measure the generators of the
stabilizer (and therefore perform fault-tolerant error correction),
and measure all members of the encoded Pauli group fault-tolerantly.
This provides step~\ref{Pauli} of the proof.

\section{Measurements and Stabilizers}

In~\cite{faulttol}, it proved very helpful in a number of places to
understand how the stabilizer of a state or subspace changed under
measurements.  The procedure for qubits generalizes easily to higher
dimensions.

First, recall that there is more than one way to choose generators for
a given stabilizer.  Any maximal set of independent operators in the
group will suffice.  In particular, any generator $M$ can be replaced
by $NM$ for any $N \neq M$.  Similarly, the encoded $\overline{X}$ and
$\overline{Z}$ operators are only defined up to multiplication by
elements of $S$.  If we wish to measure an operator $A \in {\cal P}$,
then the first step is to put the stabilizer and $\overline{X}$ and
$\overline{Z}$ operators in a form so that all the $\overline{X}$ and
$\overline{Z}$ operators commute with $A$ and all but one of the
generators of $S$ commutes with $A$.  We can do this because if $M \in
S$ does not commute with $A$, then $M^a N$ will commute with $A$ for
some $a$ for any $N$ that commutes with $M$ (as do all
$\overline{X}$'s, $\overline{Z}$'s, and generators of $S$).  We will
not need to consider the case where $A$ commutes with everything in the
stabilizer.

This is a useful form for the stabilizer because any operator that
commutes with $A$ is not disturbed by the measurement of $A$.
Therefore, we only need to change $M$ when $A$ is measured.  Since $A
\in {\cal P}$, the possible measurement results are $\omega^a$ for $a =
0, \ldots, d-1$.  These result $\omega^a$ corresponds to applying the
projection operator 
\begin{equation}
P_a = \left( I + \omega^{-a}A + \omega^{-2a} A^2 + \cdots + \omega^{-(d-1)a}
A^{d-1} \right)/d
\end{equation}
to the state.  Assume now that
\begin{equation}
M A = \omega A M
\end{equation}
(note that when $d$ is prime, this will always be true for some power
of $M$).  Then
\begin{equation}
M P_a M^\dagger = \frac{1}{d} \sum_{j=0}^{d-1} \omega^{-ja} M A^j
M^\dagger
= \frac{1}{d} \sum_{j=0}^{d-1} \omega^{-j(a-1)} A^j
= P_{a-1} \enspace.
\end{equation}
Thus, if we measure $A$ and get the result $a$, by applying $M^a$ we
can produce the same state we would have gotten if we had gotten the
result $0$.  I will assume below that any measurement is followed by
such a correction.  Once this correction is performed, the new state
is a $+1$ eigenvector of $A$, so $A$ should be added to the
stabilizer.  It is not an eigenvector of $M$, so $M$ is removed from
the stabilizer.  All of the other generators (which have been put in a
form where they commute with $A$) are unchanged.

\section{Gates Derived from SUM}

Suppose we have the ability to perform the SUM gate between any pair
of qudits in our computer, as well as the ability to perform the Pauli
group and to measure operators in the Pauli group.  I will now show
that we can apply the full Clifford group to the computer.

Suppose we consider a single unknown qudit and prepare a second
ancilla qudit in the state $|0\rangle$.  This two-qudit system can be
described by the stabilizer $I \otimes Z$.  The logical Pauli group is
generated by $\overline{X} = X \otimes I$ and $\overline{Z} = Z
\otimes I$.  Now perform a SUM gate from the first qudit to the second
qudit.  The stabilizer is $Z^{-1} \otimes Z$, $\overline{X} = X
\otimes X$, and $\overline{Z} = Z \otimes I$.

Suppose we were now to measure the operator $A = I \otimes XZ$.  Then
$M = Z^{-1} \otimes Z \in S$ and $M A = \omega A M$.  Therefore, this
measurement results in the stabilizer $I \otimes XZ$, and $\overline{X}
= X Z^{-1} \otimes X Z$ and $\overline{Z} = Z \otimes I$.  We can
discard the second qudit, and the effective transformation on the
first qudit is
\begin{eqnarray}
X & \rightarrow & X Z^{-1} \enspace, \\
Z & \rightarrow & Z \enspace.
\end{eqnarray}
This is the gate $P^{-1}$. $d-1$ iterations of it will produce the $P$
gate.

Alternatively, we could have prepared the ancilla qudit so that the
stabilizer of the system began as $I \otimes X$, then performed the
SUM gate from the second qudit to the first.  The stabilizer would
then be $X \otimes X$, and $\overline{X} = X \otimes I$ and
$\overline{Z} = Z \otimes Z^{-1}$.  Then we measure $A = I \otimes X
Z^{-1}$ and choose $M = X \otimes X$ so that $M A = \omega A M$.  The
final stabilizer is $I \otimes X Z^{-1}$, so we discard the second qudit,
leaving
\begin{eqnarray}
X & \rightarrow & X \enspace, \\
Z & \rightarrow & X Z \enspace.
\end{eqnarray}
Call this gate $Q$.  Then $R^{-1} = X Q P^{-1} Q$, and $R = R^{-3}$.

Now suppose we again prepare the ancilla in the $+1$ eigenstate of
$X$, but now perform $s$ SUM gates from the second qudit to the first
instead of one.  The stabilizer is $X^s \otimes X$, $\overline{X} = X
\otimes I$, and $\overline{Z} = Z \otimes Z^{-s}$.  This time we
measure $A = Z \otimes I$.  This results in stabilizer $Z \otimes I$,
\begin{equation}
\overline{X} = \left( X \otimes I \right) \left(X^s \otimes
X \right)^{-s^{-1}} = I \otimes X^{-s^{-1}} \enspace,
\end{equation}
and $\overline{Z} = Z \otimes Z^{-s}$.  Therefore, discarding the
first qudit leaves the transformation
\begin{eqnarray}
X & \rightarrow & X^{-s^{-1}} \enspace, \\
Z & \rightarrow & Z^{-s} \enspace.
\end{eqnarray}
By choosing an appropriate $s$, we can therefore perform an arbitrary
$S$ gate.  Note that in this case, the data ends up in what was
originally the ancilla qudit.

I have shown how to produce the $P$, $R$, and $S$ gates from the SUM
gate.  Therefore, given the SUM gate, we can produce the full Clifford
group.  This completes step~\ref{Clifford} of the proof.

\section{Producing the SUM gate for any stabilizer code}
\label{constructClifford}

To see how to construct the SUM gate between any pair of encoded
qudits, first consider two unencoded qudits.  Introduce a third qudit
in the state $|0\rangle$.  The stabilizer at this point is $I \otimes
I \otimes Z$.  Assume we can do Pauli group measurements, even
entangled ones, and perform operators in the Pauli group.  Let us
first measure the operator $I \otimes X \otimes X^{-1}$.  This becomes the
stabilizer.  The logical Pauli group generators are
\begin{eqnarray}
\overline{X}_1 & = & X \otimes I \otimes I \enspace, \\
\overline{X}_2 & = & I \otimes X \otimes I \enspace, \\
\overline{Z}_1 & = & Z \otimes I \otimes I \enspace, \\
\overline{Z}_2 & = & I \otimes Z \otimes Z \enspace.
\end{eqnarray}

Now measure $Z \otimes I \otimes Z$.  It becomes the new stabilizer,
and
\begin{eqnarray}
\overline{X}_1 & = & X \otimes X \otimes X^{-1} \enspace, \\
\overline{X}_2 & = & I \otimes X \otimes I \enspace, \\
\overline{Z}_1 & = & Z \otimes I \otimes I \enspace, \\
\overline{Z}_2 & = & I \otimes Z \otimes Z \enspace.
\end{eqnarray}
Finally, measure $I \otimes I \otimes X$ and discard the last qudit.
This leaves us with
\begin{eqnarray}
\overline{X}_1 & = & X \otimes X \enspace, \\
\overline{X}_2 & = & I \otimes X \enspace, \\
\overline{Z}_1 & = & Z \otimes I \enspace, \\
\overline{Z}_2 & = & Z^{-1} \otimes Z \enspace.
\end{eqnarray}
This we recognize as the transformation induced by the SUM gate, so
this series of entangled measurements has performed the SUM gate
between these two qudits.

Now, to apply this to a quantum code, we just need to be able to
measure entangled logical Pauli group operators between any pair of
encoded qudits.  If the qudits are in the same block, this is
straightforward.  For instance, if they are in slots $i$ and $j$, the
encoded version of $X \otimes X$ is just $\overline{X}_i
\overline{X}_j$.  This is in the Pauli group too, so we know how to
measure it.

If the qudits are in different blocks, it is not much harder.  Instead
of using an $a$-qudit ancilla state, we use an $(a+b)$-qudit ancilla
state (where $a$ and $b$ are the weights of the operators
$\overline{X}_i$ and $\overline{X}_j$), which is again in the
superposition of all states whose registers sum to $0 \bmod d$.  The
operator we wish to measure is $\overline{X}_i \otimes
\overline{X}_j$, which is in the Pauli group.  By performing the
appropriate transversal Clifford group operation, we rotate this to be
the tensor product of $Z$'s and perform SUM gates from the appropriate
qudits to the corresponding ancilla qudits, then perform the inverse
Clifford group operator to restore the state to its original form.
Then we measure the $a+b$ ancilla qudits, and this tells us the
eigenvalue of the measured operator.  We use an $(a+b)$-qudit ancilla
instead of an $a$-qudit ancilla plus a $b$-qudit ancilla because we do
not wish to be able to find the eigenvalues of $\overline{X}_i \otimes
I$ and $I \otimes \overline{X}_j$ separately, only their product.

Therefore, given an encoded ancilla qudit which is initialized to
$|0\rangle$, by performing the encoded version of the above entangled
measurements, we can perform a SUM gate between any pair of encoded
qudits anywhere in the computer.  Note that the ancilla qudit can
itself be anywhere in the computer; it need not be in the same block
as either data qudit, or in the corresponding place in a different block.

Given the SUM gate and the results of the previous section, we can
perform the full Clifford group on the encoded data for any stabilizer
code.  This completes the proof up to step~\ref{move}.  This part of
the proof is a significant improvement on the method used
in~\cite{faulttol}.  In that paper, it was necessary to introduce full
ancilla blocks to perform a CNOT.  Here, we need only a single logical
ancilla qudit.  In the case where a block may encode many qudits, this
can be a major improvement.  The price is that we must potentially perform
entangled measurements on more than one block.  This means we will
have to use larger ancilla states for the measurement; this results in
a greater potential for error, so we will have to repeat the
measurement more times, and perhaps perform error correction a bit
more often.  However, in many situations, the total number of physical
gates we need will decrease.

Note that this procedure works just as well if the two logical qudits
involved in the SUM are in blocks made up of different numbers of
physical qudits.  This means we can interact qudits encoded with
different sorts of codes fault-tolerantly, or change the encoding of a
single qudit without losing the protection against errors at any time.

\section{Completing the Universal Set of Gates}

The set of universal gates can be completed by adding the
higher-dimensional analog of the Toffoli gate~\cite{universal}
\begin{equation}
|a\rangle |b\rangle |c\rangle \rightarrow |a\rangle |b\rangle |c +
ab\rangle \enspace.
\end{equation}
It turns out that a generalization of Shor's fault-tolerant
construction of the Toffoli gate~\cite{shorFT} will work here. 

Suppose we prepare a three-qudit ancilla in the $+1$ eigenstate of the
three operators
\begin{eqnarray}
M_1 & = & \left(X \otimes I \otimes I \right) {\rm SUM} (2 \rightarrow 3)
\enspace, \\
M_2 & = & \left(I \otimes X \otimes I \right) {\rm SUM} (1 \rightarrow 3)
\enspace, \\
M_3 & = & \left(I \otimes I \otimes Z \right) {\rm PHASE} (1, 2)^{-1}
\enspace.
\end{eqnarray}
${\rm SUM} (i \rightarrow j)$ is a SUM gate performed with the $i$th
qudit as control and the $j$th qudit as target.  ${\rm PHASE} (i, j)$
is the PHASE gate
\begin{equation}
{\rm PHASE} |a\rangle |b\rangle = \omega^{ab} |a\rangle |b\rangle
\end{equation}
performed on the $i$th and $j$th qudits.  One important fact to note
is that both of these gates are in the Clifford group.  Since we have
already constructed the Clifford group, this will enable us to also
construct the Toffoli gate.  The appropriate state is
\begin{equation}
|A\rangle = \sum_{a,b} |a\rangle |b\rangle |ab\rangle \enspace.
\end{equation}

Now, given three data qudits, perform inverse SUM gates (i.e.,
$|a\rangle|b\rangle \rightarrow |a\rangle|b-a\rangle$) from the first
and second ancilla qudits to the first and second data qudits,
respectively, and a SUM gate from the third data qudit to the third
ancilla qudit.  Now we measure the last three qudits, the original
data qudits, in the bases $Z$, $Z$, and $X$, respectively.  After
performing the appropriate corrections (which will likely involve
gates from the Clifford group as well as the Pauli group), we are left
with the data in the first three qudits, which were originally the
ancilla qudits.  It turns out that after these operations, a Toffoli
gate has been performed on the data qudits.

To construct the appropriate ancilla state $|A\rangle$, we can again
follow Shor.  The various states
\begin{equation}
|A_j\rangle = \sum_{a,b} |a\rangle |b\rangle |ab + j\rangle
\end{equation}
for $j=0, \ldots, d-1$ are related to $|A\rangle$ by
\begin{equation}
|A_j\rangle = \left(I \otimes I \otimes X^j \right) |A\rangle
\enspace.
\end{equation}
The states $|A_j\rangle$, like $|A\rangle$, are $+1$ eigenstates of
$M_1$ and $M_2$, but $M_3 |A_j\rangle = \omega^j |A_j\rangle$.

Furthermore, note that
\begin{equation}
\sum_{j=0}^{d-1} |A_j\rangle = \sum_{a,b,c} |a\rangle |b\rangle
|c\rangle \enspace.
\end{equation}
This last state is easily constructed as the Fourier transform of
$|000\rangle$.  Then by measuring $M_3$ for this state, we can
collapse the state into one of the states $|A_j\rangle$.  By
applying the operator $X^{-j}$, we get the state $|A\rangle$.

To measure $M_3$ fault-tolerantly for a quantum code, we prepare the
CAT state $\sum_j |jj \cdots j \rangle$ (using the same number of
qudits as in a block of the code).  $M_3$ is in the Clifford group, so
we can perform it by some sequence of tranversal operations and
measurements.  By conditioning the appropriate operations for the
$i$th qudit on the $i$th qudit of the CAT state, we can conditionally
perform $M_3$ on the code depending on the CAT state.  Note that a
conditional operation in the $d$-dimensional case means applying
$M_3^j$ when the control qudit is in the state $|j\rangle$.  Once we
have done this, when the code is in an eigenstate of $M_3$ with
eigenvalue $\omega^s$, the CAT state ends up in the state
\begin{equation}
|CAT_s\rangle = \sum_j \omega^{js} |jj \cdots j \rangle \enspace.
\end{equation}
The various states $|CAT_s\rangle$ are orthogonal to each other, and
so can be distinguished by an appropriate measurement.  This therefore
gives us a measurement of $M_3$, completing the construction of the
Toffoli gate and step~\ref{universal} of the proof.

Because everything we do is transversal, single qudit errors in the
CAT state cannot become more than single qudit errors in any single
block of the code.  Naturally, after creating the CAT state we should
verify it to make sure there are no correlated errors.  In addition, a
single qudit error in the CAT state could give us the wrong
measurement result.  Therefore, the measurement of $M_3$ should be
repeated in order to sufficiently increase our confidence in the result.

\section*{Acknowledgements}

I would like to thank Manny Knill and Raymond Laflamme for helpful
discussions, and Michael Nielsen for suggesting the name ``Pauli group.''
This paper is preprint LA-UR 98-270.

\end{document}